\documentclass[10pt]{sig-alternate-10pt}

\usepackage{epsfig,endnotes}

\usepackage{times}  
\usepackage{hyperref}

\usepackage{amsmath,amssymb}
\usepackage{url}
\usepackage{multirow}
\usepackage[usenames]{color}
\usepackage{balance}
\usepackage[hyphenbreaks]{breakurl}
\usepackage{algorithm}
\usepackage{algorithmic}

\usepackage{subfigure}

\usepackage{array}
\newcolumntype{L}[1]{>{\raggedright\let\newline\\\arraybackslash\hspace{0pt}}m{#1}}
\newcolumntype{C}[1]{>{\centering\let\newline\\\arraybackslash\hspace{0pt}}m{#1}}
\newcolumntype{R}[1]{>{\raggedleft\let\newline\\\arraybackslash\hspace{0pt}}m{#1}}

\newcommand{\para}[1]{{\vspace{1pt} \bf \noindent #1 \hspace{6pt}}}

\newcommand\yedit[1]{{\color{black} #1}}
\newcommand\jedit[1]{{\color{black} #1}}
\newcommand{\eg}{{\em e.g.,\ }}
\newcommand{\ie}{{\em i.e.\ }}

\newcommand{\secspacesm}{\vspace{-0.08in}}

\newenvironment{packed_itemize}{
\begin{list}{\labelitemi}{\leftmargin=1.em}
  \setlength{\itemsep}{2pt}
  \setlength{\parskip}{0pt}
  \setlength{\parsep}{0pt}
  \setlength{\headsep}{0pt}
  \setlength{\topskip}{0pt}
  \setlength{\topmargin}{0pt}
  \setlength{\topsep}{0pt}
  \setlength{\partopsep}{0pt}  
}{\end{list}}

\newfont{\mycrnotice}{ptmr8t at 7pt}
\newfont{\myconfname}{ptmri8t at 7pt}
\let\crnotice\mycrnotice%
\let\confname\myconfname%

\begin{document}

\title{Wireless Side-Lobe Eavesdropping Attacks}

\author{
\alignauthor Yanzi Zhu, Ying Ju$^{\S\dag}$, Bolun Wang, 
Jenna Cryan$^\ddag$, Ben Y. Zhao$^\ddag$, Haitao Zheng$^\ddag$  \\
\affaddr{University of California, Santa Barbara} \;
\affaddr{$^\S$Xi'an Jiaotong University}\\
\affaddr{$^\dag$State Radio Monitoring Center} \; 
\affaddr{$^\ddag$University of Chicago} \\
\affaddr{\{yanzi, bolunwang\}@cs.ucsb.edu, juyingtju@163.com,} \\
\affaddr{\{jennacryan, ravenben, htzheng\}@cs.uchicago.edu}
}

\maketitle

 \vspace{-0.2in}
\begin{abstract}

Millimeter-wave wireless networks offer high throughput and can
(ideally) prevent eavesdropping attacks using narrow, directional beams.
\yedit{
Unfortunately, imperfections in physical hardware mean today's antenna arrays
all exhibit side lobes, signals that carry the same sensitive data 
as the main lobe.}
Our work presents results of the first
experimental study of the security properties of mmWave transmissions against
side-lobe eavesdropping attacks. We show that these attacks on mmWave links are
highly effective in both indoor and outdoor settings, 
and they cannot be eliminated by improved hardware or currently 
proposed defenses.
\end{abstract}

 \vspace{-0.06in}
\section{Introduction}
\label{sec:intro}

Wireless communication has always been more vulnerable to attacks than its
wired counterparts. The fact that wireless signals are broadcast means they are
more easily eavesdropped. This weakness has been
exploited in many wireless networks~\cite{crack_bluetooth,
  sheldon2012insecurity, crack_wepwpa}. Even more recent security protocols
like WPA2-PSK have been successfully compromised by snooping attacks~\cite{crack_wpa2psk, nakhila2015} via simple tools~\cite{aircrack}.
\yedit{Despite existing encryptions,
one can still infer the specific sources of traffic by observing just packet sizes
and counts in data transmissions~\cite{pet-http, marc06inferhttp}.}

\yedit{While we continue to improve encryption algorithms,
an equally promising direction} 
is to use wireless beamforming to defend against
eavesdroppers at the physical layer. Beamforming allows a transmitter (TX) to
send a highly focused, directional signal towards a target receiver (RX), so
that nearby attackers not directly between the two endpoints cannot capture
the transmission.  The narrow beam is built by leveraging signal
cancellations among multiple antennas in a phased array\footnote{ We do not
  consider horn antennas as they are bulky, expensive, and can only be
  rotated mechanically. They are not suitable for our application scenarios.}, and is most easily
built on \textit{millimeter-wave} (mmWave) transmitters~\cite{mmwave_secure}. 
For example, 60GHz phased arrays could fit on small devices like smartphones, 
and can generate highly focused beams 
(\eg 3$^\circ$ using 32$\times$32 antennas) while achieving Gbps throughput.

While earlier applications focused on short-range indoor applications, \eg
home routers~\cite{tplink11ad} and wireless virtual
reality headsets~\cite{htcvive60g}, new applications of mmWave leverage its
high directionality and throughput for long-range communication.  Many such
applications have already been deployed.  Facebook has deployed a mesh 
network using 60GHz communications in downtown San
Jose~\cite{facebook_sanjose}.  Google is considering  replacing wired fiber
with mmWave to reduce cost~\cite{googlefiber}. Academics have proposed
picocell networks using mmWave signals towards next 5G
network~\cite{marzi_globecom15, picocell_tracking, zhu14a}.

\begin{figure*}[ht]
    \begin{minipage}{0.29\textwidth}
        \vspace{-0.08in}
        \centering
        \includegraphics[width=1\textwidth]{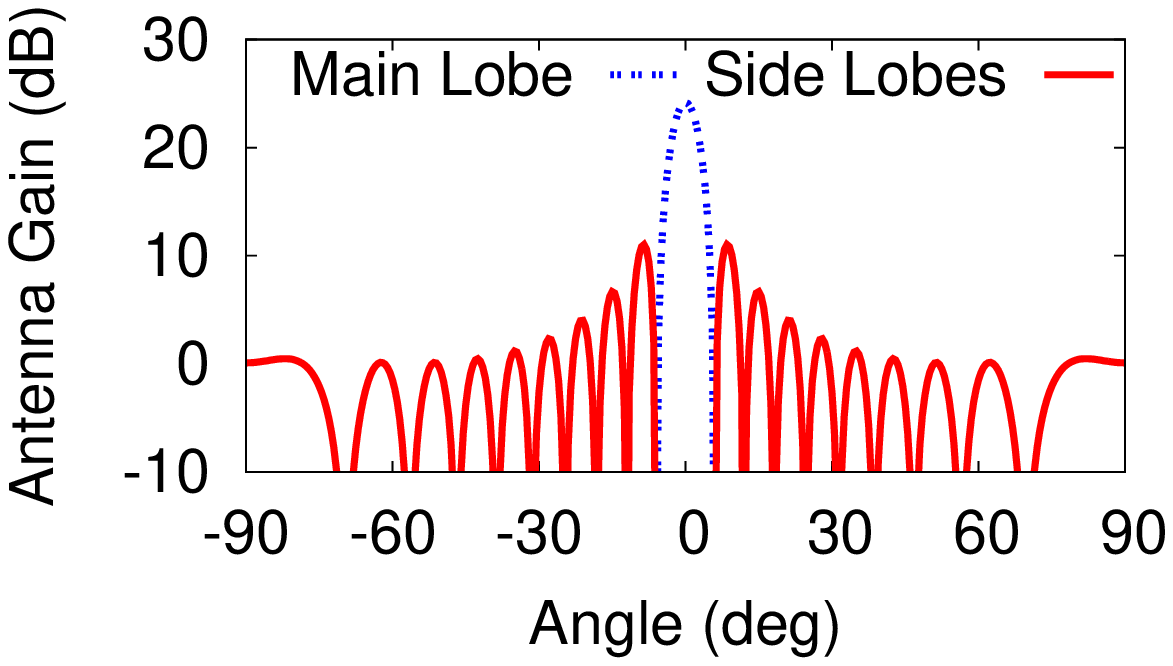}
        \vspace{-0.28in}
        \caption{Example of side lobes of a 16$\times$8 array (horizontal plane).}
        \label{fig:sidelobe_example}
    \end{minipage}
    \hfill
    \begin{minipage}{0.685\textwidth}
        \centering
        \includegraphics[width=1\textwidth]{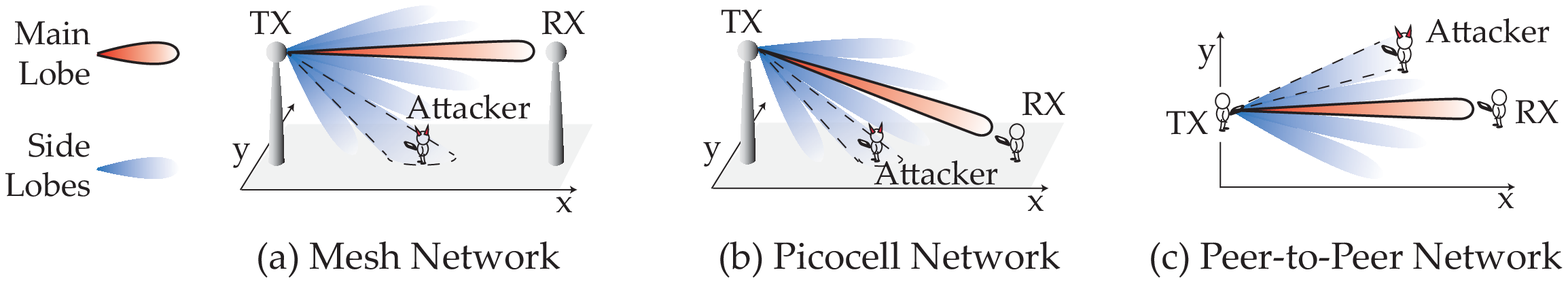}
        \vspace{-0.22in}
        \caption{Illustration of three application scenarios we test the 
        eavesdropping attack. Attacker could eavesdrop through side lobes (blue) 
        to decode information transmitted in the main lobe (red).}
        \label{fig:scenarios}
    \end{minipage}
    \vspace{-0.1in}
\end{figure*}

With a growing number of deployed networks and applications, understanding
physical properties of mmWave is critical. One under-studied
aspect of directional transmissions is the artifact of array {\em side
lobes}. Fig.~\ref{fig:sidelobe_example} shows an example 
of the series of side lobes pointing in different directions. 
Side lobes are results of imperfect signal cancellation among antenna
elements. While weaker than the main lobe, side lobes carry the same
information, and can be exploited by eavesdroppers to recover the
transmission.  As physical imperfections, they are very difficult to eliminate.

 In this paper, we conduct the first empirical study of the security
 properties of mmWave communications against side-lobe eavesdropping attacks.
\yedit{While theoretical studies have shown the problem of side-lobe leakage~\cite{kim2017analysis}, it is never validated using network measurements, especially for long-range 
communications.}
We use a commercial 60GHz testbed from Facebook's Terragraph project~\cite{terragraph} to evaluate the
effectiveness of side-lobe eavesdropping in both indoor and outdoor
scenarios. Specifically, we answer three key questions:

\begin{packed_itemize}
 \vspace{-0.06in}
\item \para{How severe is mmWave side-lobe eavesdropping?
    (\S\ref{sec:measurement})} We observe that side-lobe eavesdropping is
  incredibly effective in both indoor and outdoor scenarios.  Attacker
  can recover transmission in a large area with high success rate (details
  below).  \yedit{Particularly for outdoor scenarios, 
   most eavesdropping areas 
  are connected, and the attacker can move freely and launch stealthy attacks.}
    \begin{table}[h]
    \vspace{-0.08in}
    \raggedleft
    \small
    \begin{tabular}{|l|C{1.16cm}|C{1.16cm}|C{1.16cm}|}
    \hline
    \multirow{2}{*}{Eavesdropping Area ($m^2$)} & 
    \multicolumn{3}{c|}{Attacker's Packet Success Rate} \\ 
    \cline{2-4} 
     & $>$10\% & $>$50\% & $>$95\% \\ 
    \hline
    Mesh & 79 & 64.6 & 55 \\ 
    \hline
    Picocell & 109 & 88.6 & 54 \\ 
    \hline
    Peer-to-Peer & 16.6 & 15.7 & 13.1 \\ 
    \hline
    \end{tabular}
    \vspace{-0.1in}
    \end{table}

  \item \para{Can better mmWave hardware improve security?
      (\S\ref{sec:simulation})} We find that improved hardware can only
    reduce the impact of the eavesdropping attack, but not fully defend
    against it.  Eavesdropping side lobes is still possible even after
    removing hardware artifacts from antennas and deploying more antenna
    elements.

  \item \para{Are existing defenses effective against side-lobe eavesdrop
      attacks? (\S\ref{sec:existing})} Although existing defenses show
    promising results against single-device eavesdroppers, they either impose
    impractical hardware requirements, or remain vulnerable against more
    advanced attackers, \eg those with multiple devices.
\end{packed_itemize}

 \vspace{-0.15in}
\section{Background}
\label{sec:background}

To provide context for later study, we first describe
the adversarial model 
and then our measurement methodology.

\para{Adversarial Model.} We consider {\em passive} eavesdropping, where an
attacker listens to side-lobe signals and recovers packet header or payload.
 The
attacker stays hidden from its victim TX and RX, but is unable to manipulate
the communication between the victims. Without knowing the attacker's physical
location, victims cannot apply conventional defenses like
null-forming\footnote{If TX knows the attacker's location, it can change its
  radiation pattern to nullify signals towards that
  location to avoid attacks~\cite{nullforming}.}.  

We do not consider eavesdropping attacks on the main lobe of the
transmission. Such an attack would affect the 
communication between TX and RX, as the attacker has to stay inside the main
lobe or use a reflector, and thus can be
detected~\cite{steinmetzer2015eavesdropping}.
Finally, we assume the attacker has one or more synchronized devices as
powerful as the victim's hardware.  The attacker knows the victim's location
and hardware configuration\footnote{This information is often publicly
  available, or could be derived from simple techniques, \eg device
  localization.}.  The attacker and his device(s) are free to move around the
victims.

\para{Application Scenarios.}
We consider three practical scenarios where 
mmWave signals are commonly used:
mesh networks~\cite{facebook_sanjose}, picocell networks~\cite{zhu14a}, 
and indoor peer-to-peer 
transmissions~\cite{tplink11ad,htcvive60g}.
Fig.~\ref{fig:scenarios} shows an illustration of the three.

\yedit{mmWave signals are commonly considered for indoor peer-to-peer 
scenarios (Fig.~\ref{fig:scenarios}(c)), \eg 
virtual reality~\cite{htcvive60g, abari17enabling}
and wireless display~\cite{delldock}. Here TX and RX are within very short 
range ($\leq$10m) and often at the same height ($\sim$1m).}
As mmWave signals degrade much faster than lower frequency signals
in the air, it is less known that
they can also be used outdoor for long-range communications
(20--200m). For example, Facebook has deployed a mesh network 
in downtown San Jose~\cite{facebook_sanjose},
supporting up to 200m link using 60GHz phased array radios\footnote{Compared
  to horn antennas, phased arrays offer robust real-time
  link adaptation by eliminating mechanical steering.}. Researchers~\cite{marzi_globecom15, zhu14a} also propose picocell 
networks using 60GHz signals.
In both scenarios, TX is mounted higher than human height, \eg 6m. 
Depending on the scenario, RX is either mounted at a similar height or 
on the ground, shown in Fig.~\ref{fig:scenarios}(a) and (b), respectively.

\begin{figure}[t]
    \centering
    \includegraphics[width=0.28\textwidth]{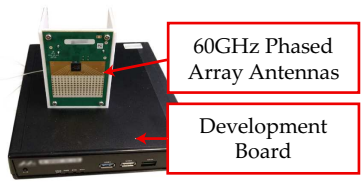}
    \vspace{-0.1in}
    \caption{Our 60GHz testbed with 16$\times$8 antenna array.}
    \label{fig:testbed}
\end{figure}

\para{Measurement Hardware.} 
Our testbed consists of three identical 60GHz radios. We use them as TX, RX, 
and the attacker. Each radio has a 16$\times$8 rectangular phased array 
(Fig.~\ref{fig:testbed}) and follows the 802.11ad single-carrier 
standard for 60GHz communication~\cite{802.11ad}. 
Our radios are designed for outdoor mesh network scenario with a 
maximum Equivalent Isotropically Radiated Power (EIRP) of 32dBm, 
supporting 1Gbps (QPSK) transmissions at 200m range (line-of-sight).
But we could re-purpose these radios for picocell and 
peer-to-peer scenarios as well, by lowering the EIRP.
Each receiving radio can report received signal-to-noise-ratio (SNR) 
of each packet in real time.

\begin{table}
\centering
\resizebox{\columnwidth}{!}{
\begin{tabular}{|l|c|c|c|c|c|c|}
\hline
\multirow{2}{*}{
Scenario
} & \multicolumn{2}{c|}{TX} & \multicolumn{3}{c|}{RX} & \multirow{2}{*}{\begin{tabular}[c]{@{}c@{}}Examined\\Area\\(m$^2$)\end{tabular}} \\ \cline{2-6}
 & \begin{tabular}[c]{@{}c@{}}EIRP \\ (dBm)\end{tabular} & \begin{tabular}[c]{@{}c@{}}Height \\ (m)\end{tabular} & \begin{tabular}[c]{@{}c@{}}Distance \\ to TX\\(m)\end{tabular} & \begin{tabular}[c]{@{}c@{}}Height \\ (m)\end{tabular} &  \begin{tabular}[c]{@{}c@{}}Max\\Throughput\\ (Gbps)\end{tabular} & \\ 
\hline
Mesh & 32 & 6 & 200 & 6 & 1.0 & 10$\times$20 \\ 
\hline
Picocell & 32 & 6 & 50 & 1 & 1.5 & 10$\times$20 \\ 
\hline
Peer-to-Peer & 23 & 1 & 10 & 1 & 1.5 & 4$\times$5 \\ 
\hline
\end{tabular}
}
\vspace{-0.1in}
\caption{Detailed experiment setup and configurations.}
\label{tab:setup}
\vspace{-0.1in}
\end{table}

\para{Measurement Setup.}
We place our testbed radios at different heights and distances apart to 
emulate the three application scenarios. In all scenarios, TX sends 
32KB TCP packets to RX at 1Gbps by default.
Equipment placement details and specifications are listed in 
Table~\ref{tab:setup}.
In particular for (c) peer-to-peer,
we choose 23dBm EIRP the same as the commodity 60GHz chipset from
Wilocity~\protect\cite{wilocity}. Given TX's EIRP and 
the distance from victim RX to TX, RX can at best communicate 
with TX at 1Gbps, 
1.5Gbps, and 1.5Gbps with less than 5\% packet loss
in mesh, picocell, and peer-to-peer networks, 
respectively. Further reducing TX power will affect RX's performance.

During transmission, we move the attacker radio around TX to eavesdrop 
side lobes at different locations. We grid the area around TX 
(200$m^2$ for two outdoor scenarios and 20$m^2$ for the indoor scenario) 
into 816 (34$\times$24) rectangles. In each grid, we face the attacker 
radio at TX and 
eavesdrop the transmission for 30$s$. Our testbed could record 100k packet 
samples and 30 SNR values in each grid. In each application 
scenario, we collected a total of 80 million packets and 24k SNR measurements.

 \vspace{-0.1in}
\section{Effectiveness of Eavesdropping}
\label{sec:measurement}

From our collected measurements, we now present the severity of side-lobe
eavesdropping under three mmWave network scenarios.
We use the following two metrics to quantify 
the effectiveness of side-lobe eavesdropping.

\begin{packed_itemize} \vspace{-0.1in}
    \item {\em Packet success rate (PSR)}
    measures the percentage of packets the attacker could successfully 
    retrieve from eavesdropping through side lobes, calculated from 
    100k packets per location.  When the attacker's PSR
    is no less than that of the victim RX ($>$95\% in our
    experiments), we consider it to be a  {\em full} attack.

    \item {\em Eavesdropping area} 
    measures the area where the attacker can achieve PSR higher than a given
    threshold by eavesdropping on side lobe signals. 
\end{packed_itemize}
 \vspace{-0.06in}

\begin{figure*}[t]
  \centering
  \mbox{
    \includegraphics[width=0.33\textwidth]{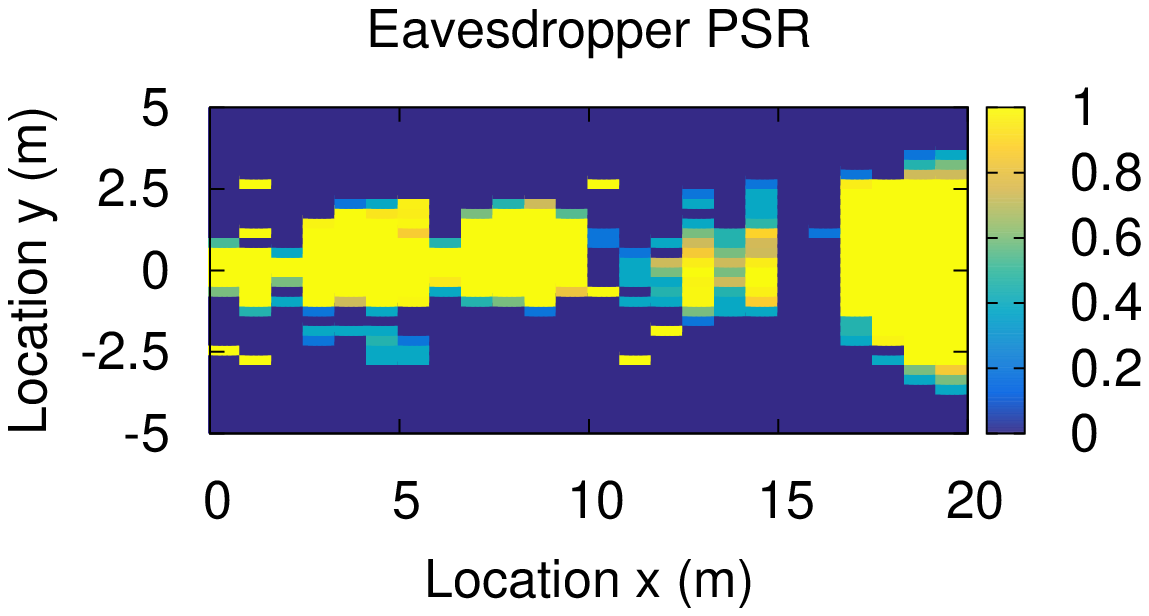}
    \hfill
    \includegraphics[width=0.33\textwidth]{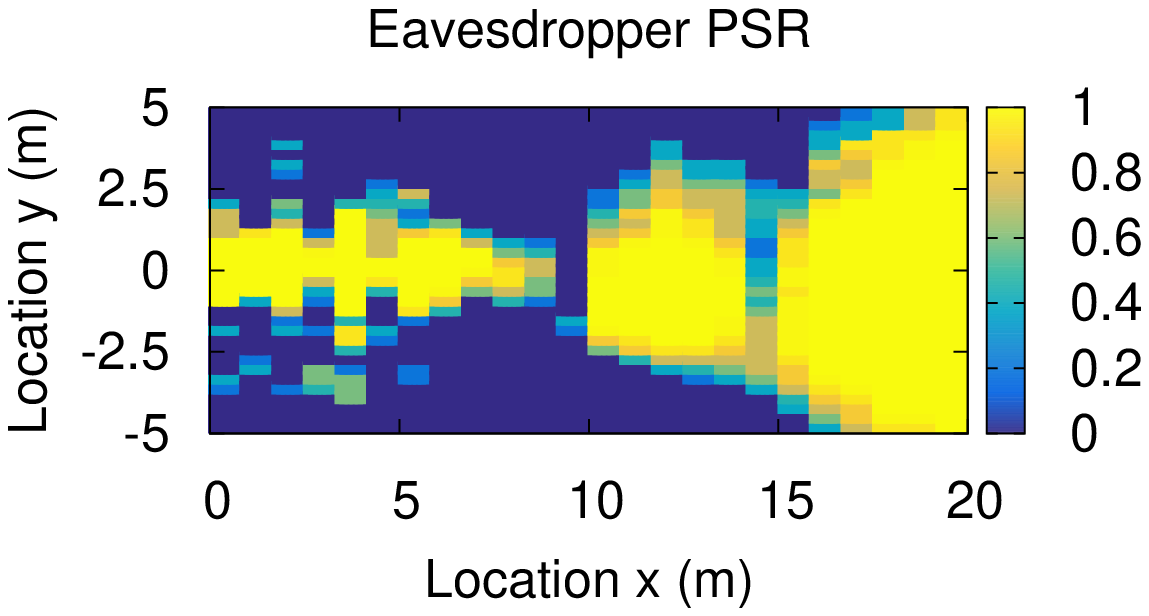}
    \hfill
    \includegraphics[width=0.33\textwidth]{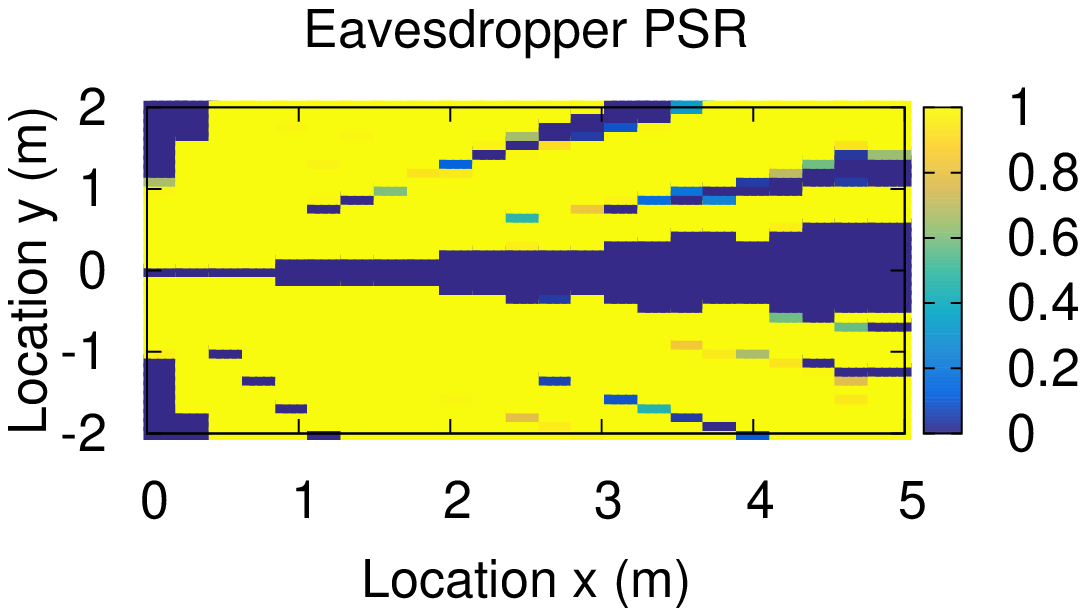}
  }
  \mbox{
    \subfigure[Mesh Network]{
      \includegraphics[width=0.315\textwidth]{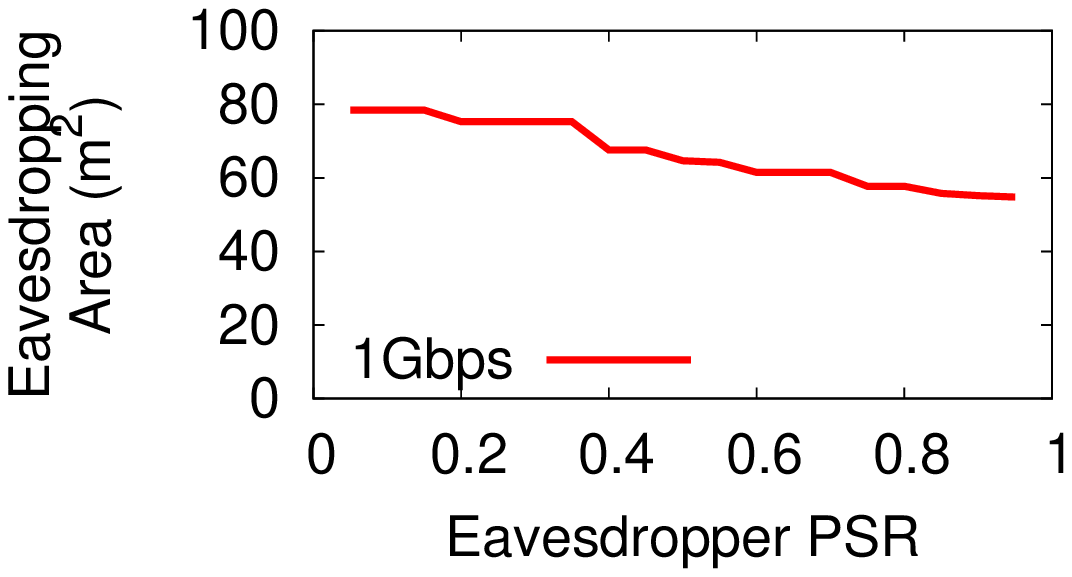}
      \label{fig:sidelobe_mesh}
    }
    \hspace{0.01in}
    \subfigure[Picocell Network]{
      \includegraphics[width=0.315\textwidth]{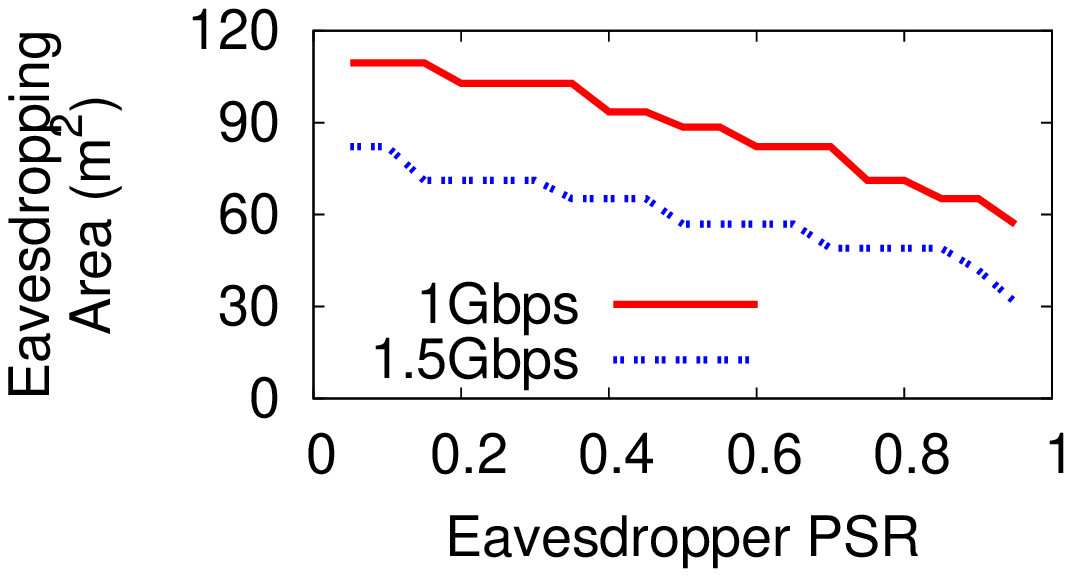}
      \label{fig:sidelobe_pico}
    }
    \hspace{0.01in}
    \subfigure[Peer-to-Peer Network]{
      \includegraphics[width=0.315\textwidth]{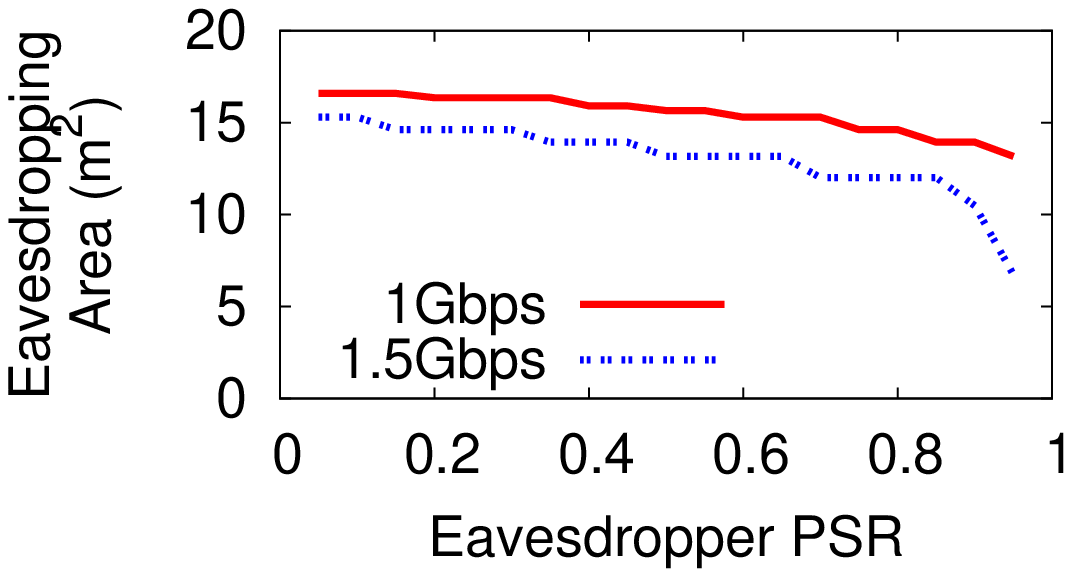}
      \label{fig:sidelobe_indoor}
    }
  }
  \vspace{-0.1in}
  \caption{Effectiveness of side-lobe eavesdropping under three mmWave
    scenarios.
    For each scenario, the top plot shows attacker's packet success rate (PSR) at 1Gbps at different \yedit{side-lobe} locations. 
    TX is at (0, 0) and beams towards
    RX along x-axis. The bottom one shows how the eavesdropping area changes with PSR thresholds at different link rates. 
  }
  \label{fig:sidelobe}
  \vspace{-0.1in}
\end{figure*}

\secspacesm
\subsection{Mesh Network}

We begin by showing the effectiveness of eavesdropping in an outdoor mesh network.
During transmission, the main lobe points towards RX and side lobes point 
towards the ground. The eavesdropper moves freely on the ground 
and searches for locations where he could hear side-lobe signals.
Fig.~\ref{fig:sidelobe_mesh} shows the attacker's PSR at different 
locations, and how the eavesdropping area changes.

From the heatmap in the figure, 
we observe that the attack is proved very effective. In 79$m^2$ 
out of the 200$m^2$ examined area, the attacker could successfully decode
\textit{at least one} packet. 
\jedit{
Aggregated, the eavesdropping area accounts for 39.5\% of the entire area.
Large connected portions allow an attacker to act more stealthily
by moving freely through areas as large as 23$m^2$ 
rather than staying in one location.
}

Note that all eavesdropping areas center along TX's transmission 
direction (along the x-axis). This allows the attacker to easily predict 
vulnerable areas to launch attacks, because side lobes 
along the x-axis are strong enough for eavesdropping, while other side lobes 
pointing away from the x-axis suffer higher signal loss ($>$13$dB$) 
and become too weak.

We further investigated how the eavesdropping area changes given different PSR
thresholds. As shown in the lower figure in Fig.~\ref{fig:sidelobe_mesh}, the 
eavesdropping area reduces very slowly as we increase the PSR threshold.
When requiring the attacker to achieve $>$50\% PSR, the eavesdropping area 
reduces only by 19\%, down to 64$m^2$. Moreover, in a 55$m^2$ area
(69.6\% of the total eavesdropping area), the attacker could achieve 
$>$95\% PSR, the same performance achieved by RX. This further 
shows the incredible effectiveness of an eavesdropping attack in a mesh 
network scenario.

Our measurements for the mesh network cover 200$m^2$ area and already show
the severity of side-lobe eavesdropping. Although not shown in figure, we found
that more eavesdropping locations also exist outside the examined area, 
\eg $>$20$m$ away from TX. We leave further investigation of these areas 
to future work.

\subsection{Picocell Network}

Fig.~\ref{fig:sidelobe_pico} shows the eavesdropping results in a picocell network
scenario.
Similar to the mesh network scenario, the attacker could successfully 
eavesdrop the transmission in a large area. Within 109$m^2$, the attacker
could decode at least one packet, which is 54.5\% of the entire examined area.
An area of 54$m^2$ within this 109$m^2$ allows the attacker to eavesdrop 
with $>$95\% PSR, thus fully recovering the victim's transmission.
The ratio of
eavesdropping area to the entire examined area is comparable to the mesh
network scenario, which indicates similar levels of effectiveness of the 
eavesdropping attack.

Interestingly, in both the mesh and picocell networks,
the area of connected eavesdropping locations
grows larger as the attacker moves away from TX. 
This seems counter-intuitive, since signals become weaker
at farther distances due to propagation loss. However, 
in outdoor scenarios, the projection of side lobes on
the ground grows larger at farther distances. Despite the
propagation loss, the side lobes remain strong enough 
for the attacker to successfully decode the transmission, given
the sufficiently high TX power for transmissions at distances over 
100$m$. This finding appears more obvious in the picocell network
because TX's beams point downwards, causing less 
propagation loss through side lobes.

\para{Increasing link rate reduces eavesdropping area.}
Different from the mesh network, where RX remains stationary and 
achieves at most 1Gbps throughput, the victim RX in picocell is 
mobile. As RX moves closer to TX, RX could achieve higher SNR and increase 
data rate up to 1.5Gbps, while maintaining $>$95\% PSR. 
We re-configured the testbed to transmit at 1.5Gbps, and measured the 
corresponding PSR at different locations. 
The lower figure in Fig.~\ref{fig:sidelobe_pico} shows a smaller eavesdropping area
when TX increases data rate from 1Gbps to 1.5Gbps. On average, 
this reduces the eavesdropping area by 24$m^2$. 
In particular, when requiring the 
attacker to have $>$95\%, increasing throughput reduces the 
eavesdropping area from 54$m^2$ to 31$m^2$.
\yedit{The area reduces because  increasing the legit transmission rate also raises the channel quality requirement at the eavesdropper, thus
mitigating the attack to some extent.}
Yet it does not fully solve the problem, as the attacker could still 
successfully decode packets in a large area.

\vspace{-0.05in}
\subsection{Peer-to-Peer Network}

Fig.~\ref{fig:sidelobe_indoor} shows the eavesdropping performance in 
a peer-to-peer scenario. 
In a connected area of 16.6$m^2$ the attacker could decode at least one packet.
The area is significantly large (83\%),
compared to the 20$m^2$ total examined area.
When requiring $>$95\% PSR, the attacker could still decode the transmission
in  65\% (13.1$m^2$) of the total area.

Similar to the picocell scenario, both RX and TX can move freely, 
causing different distances between RX and TX. 
This allows higher SNR and higher link rate 
without degrading RX's PSR, but again, it cannot remove the eavesdropping
area completely. Still,
in an area of 7$m^2$, the attacker could decode transmissions with $>$95\% PSR.

Note that the shape of eavesdropping area in the peer-to-peer scenario differs
from those in the other two scenarios. 
This is mainly because TX sits at a much 
lower height than the other two scenarios. The attacker 
resides on the same plane of TX and RX, and captures the side-lobe
signals on the horizontal plane. As such, the eavesdropping area
follows a similar shape of the side-lobe beam pattern 
(Fig.~\ref{fig:sidelobe_example}), rather than the circular ones
observed in mesh and picocell networks. 
This observation of different shapes within eavesdropping areas 
could better guide the attacker's predictions for 
where to launch attacks based on a targeted scenario.

Although the eavesdropping area in an indoor scenario 
accounts for a larger portion of the examined area than the 
outdoor scenarios, its absolute size is significantly smaller,
thus with less potential threat.
Moreover, 60GHz signals can hardly penetrate walls, so the eavesdropping area 
for the indoor scenario 
remains bounded by its room size, further restricting 
the attacker's mobility and effectiveness of eavesdropping. Therefore, 
side-lobe eavesdropping proves much more severe in the prior two outdoor 
scenarios.

\vspace{-0.09in}
\subsection{Summary}

In all scenarios, we find that a passive attacker could
effectively eavesdrop transmissions with very high PSR in a large area.
This shows that despite the directional nature of mmWave beamforming 
transmission, side lobes still expose a significant amount of information.
Increasing transmission rate slightly mitigates 
the threat, but cannot effectively defend against the eavesdropping attack.

\section{Impact of Radio Hardware}
\label{sec:simulation}

\begin{figure}[t]
    \centering
    \includegraphics[width=0.42\textwidth]{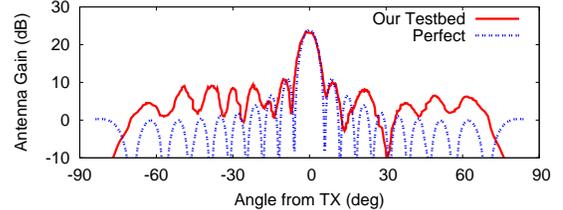}
    \vspace{-0.1in} 
    \caption{Antenna artifacts cause side lobe distortions.}
    \label{fig:sim_distortion}
    \vspace{-0.15in}
\end{figure}

\begin{figure*}[ht]
        \begin{minipage}{0.32\textwidth}
            \centering
            \includegraphics[width=0.98\textwidth]{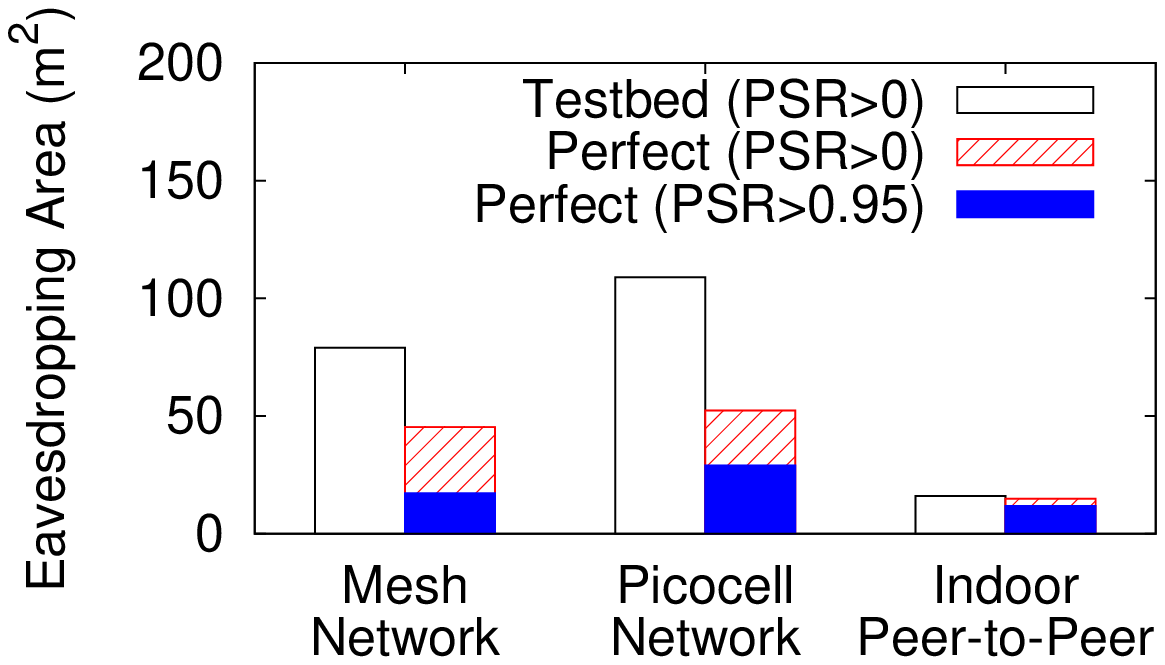}
            \vspace{-0.1in} 
            \caption{Perfect antennas 
            help mitigate eavesdropping but 
            not avoid it.}
            \label{fig:sim_perfecthw}
        \end{minipage}
        \hfill
        \begin{minipage}{0.32\textwidth}
            \centering
            \includegraphics[width=0.98\textwidth]{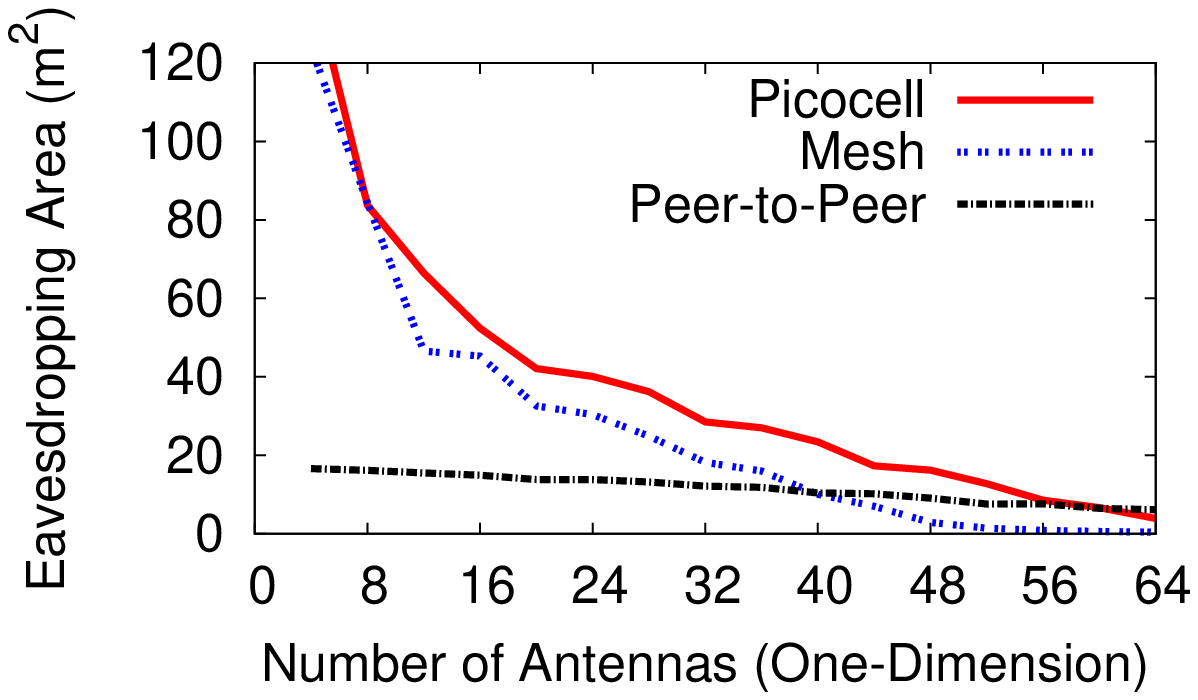}
            \vspace{-0.1in} 
            \caption{Increasing number of antennas helps reduce
            eavesdropping area.}
            \label{fig:sim_antennasize}
        \end{minipage}
        \hfill
        \begin{minipage}{0.32\textwidth}
            \centering
            \includegraphics[width=0.98\textwidth]{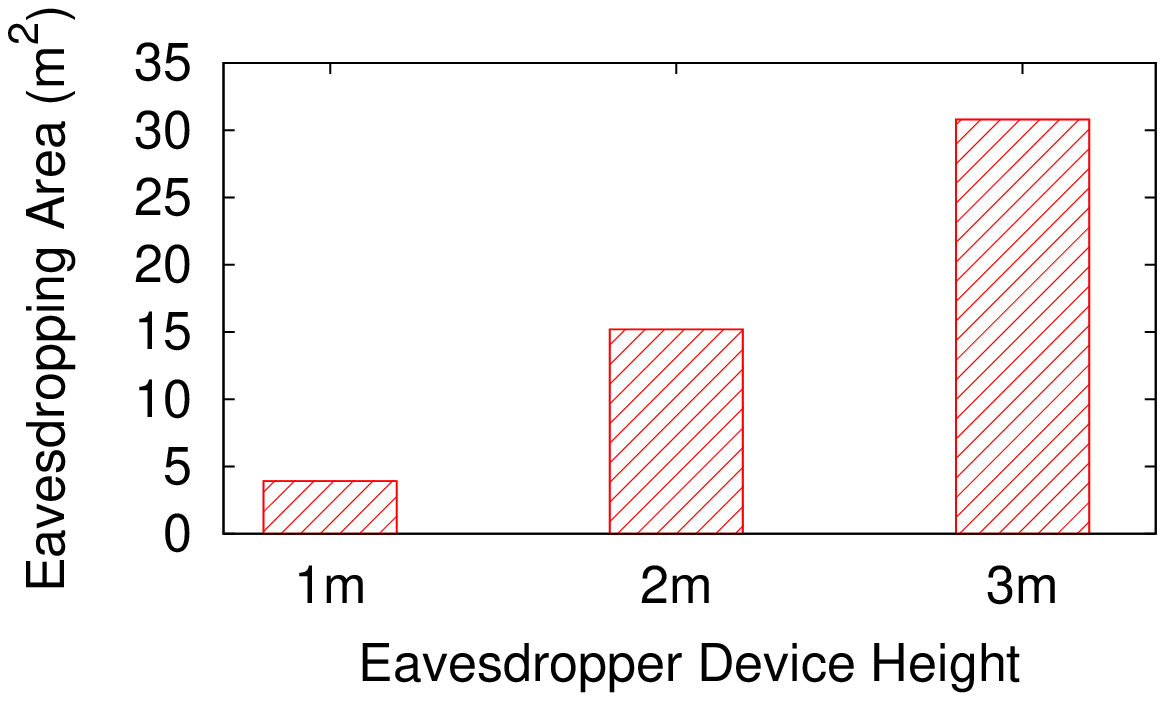}
            \vspace{-0.1in} 
            \caption{Attacker can raise device high
            to enlarge eavesdropping area.}
            \label{fig:sim_raiseheight}
        \end{minipage}
\end{figure*}

So far we have empirically measured the eavesdropping area using 
off-the-shelf 60GHz devices (16$\times$8 phased arrays). In this
section, we further explore whether upgrading array hardware can help
reduce the impact of
eavesdropping attacks.  Specifically, there are two immediate ways to improve 
mmWave array 
hardware and reduce side-lobe emissions: (1) removing implementation artifacts from the antenna 
radiation pattern,  (2) increasing the number
of antenna elements. Fig.~\ref{fig:sim_distortion} compares the ideal antenna
radiation pattern  and that of our current hardware. While the current
hardware faces distortions on side-lobe emissions, the ideal array implementation 
would produce weaker, less detectable side lobes.  Similarly, increasing the
number of antennas can also reduce the emission power of side
lobes~\cite{balanis2016antenna}, thereby reducing the performance of an eavesdropping attack.  

In the following, we study how upgrading radio hardware would reduce the
eavesdropping effectiveness. To emulate hardware configurations different
from our testbed, we used trace-driven simulations. 
Specifically, 
we apply the Friis free-space propagation model~\cite{friis} to compute an attacker's SNR 
at different locations. 
All of our testbed measurements,
along with prior works~\cite{zhu14a, zhou12}, show that
this model can accurately estimate the SNR in line-of-sight
with very small errors ($\pm 1dB$). 
At each location, we map 
simulated SNR to PSR using the empirical correlation derived from 
previous testbed experiments. We verified that this correlation 
remains stable and accurate across different application scenarios 
and link rates. Our simulations follow the same configuration 
in~\S\ref{sec:background}, with altered hardware aspects.
We also expanded the experiments by varying the height of the eavesdropping 
device and RX's locations.

\vspace{-0.05in}

\subsection{Perfect Antennas without Artifacts}

First we simulated eavesdropping attacks on three application scenarios,
using perfect antennas without artifacts. Fig.~\ref{fig:sim_perfecthw}
shows the eavesdropping areas for different scenarios, compared with our testbed
measurements. We only present results with 1Gbps, and omit 
those from other data rates as they show similar findings.

Comparing eavesdropping areas using perfect antennas and our testbed, we found
eliminating hardware artifacts reduces the eavesdropping area. 
In the mesh and picocell network scenarios, the eavesdropping area reduced
by 43\% and 52\% respectively. However, the area for the indoor peer-to-peer scenario
reduced by only 4\%, as for short-range indoor communications, 
TX's power (with 23$dBm$ EIRP)
at side lobes is high enough to allow eavesdropping.

Despite the reduced eavesdropping area, we find the remaining area is 
still large enough for attackers to move around while achieving high PSR.
In mesh, picocell, and peer-to-peer scenarios, an attacker could achieve 
full recovery of the transmission ($>$95\% PSR) in 45$m^2$, 52$m^2$, 
and 15$m^2$ respectively. Thus, 
\textbf{\textit{ removing hardware artifacts cannot fully defend against eavesdropping}}.

\vspace{-0.05in}

\subsection{Increasing Number of Antenna Elements}

In addition to removing artifacts from hardware, we increased the number
of antennas, and tested if the combination of these two techniques could
defend against the eavesdropping attacks. Fig.~\ref{fig:sim_antennasize} shows
how the eavesdropping area (with PSR $>$0) changes as we increase the 
number of antennas in the horizontal plane (our testbed uses 16
antennas in this plane). We find that in all our application scenarios, 
eavesdropping area decreases monotonically as we add more antennas. 
For example, in the picocell network scenario, using 64 antennas 
(compared to 16 in our testbed) effectively reduces the eavesdropping
area from 52.39$m^2$ down to 3.91$m^2$.
This confirms the theory that more antenna elements reduce 
side lobes' beam width and emission power, resulting in shrinking the area 
where an attacker could receive the side-lobe signals.

As well as incurring larger hardware
implementation cost and size, 
\textbf{\textit{increasing the number of antennas does not fully prevent 
an eavesdropping attack}}. For instance, in both mesh and picocell 
scenarios, a simple yet effective method for attacker is to raise 
the eavesdropping device to get closer to TX and receive stronger signals. 
This results in higher SNR
than eavesdropping on the ground, and the attacker could achieve better 
eavesdropping results. Fig.~\ref{fig:sim_raiseheight} shows its effect
in the picocell scenario. Even though TX uses 64 perfect
antennas (in the horizontal plane), an attacker could increase the eavesdropping area from 
3.91$m^2$ to 15.2$m^2$ by moving the device from 
in-hand position (1$m$) to above-head (2$m$). If attacker uses drones to
further raise the device height, the eavesdropping area
increases to 30.8$m^2$. We observed similar improvement in mesh
networks. As such, even after reconfiguring hardware with significant cost, 
an attacker could still successfully eavesdrop in large area. 
This poses a serious security threat as simple methods, like holding the device higher,
allow attackers to advance beyond hardware upgrades.
So we need new defense mechanisms.

\section{Analysis of Existing Defenses}
\label{sec:existing}

Existing defenses  
focus on adding \textit{artificial noises} to 
side-lobe signals to prevent attackers from decoding 
useful information~\cite{alotaibi_tcom16, heath_enhancing16,
juying17,ramadan_icc16, heath_tcom13, mckay13, heath_adhoc16}.
They fall under two categories, depending on 
how the noise is generated: (1) antenna-based defenses and 
(2) RF-chain-based defenses\footnote{
An RF (radio-frequency) chain refers to a set of physical
hardware components for wireless signal processing, bridging
between the antenna array and radio baseband.
}.
In this section, we analyze these defenses to study whether they
are practical and effective defending against side-lobe eavesdropping.
We summarize them in Table~\ref{table:defense}.

\para{Antenna-Based Defenses.}
This defense creates noisy side lobes by changing the radiated
signals from a subset of antenna elements. During transmission, 
TX either disables~\cite{alotaibi_tcom16, heath_tcom13}
or flips the phase~\cite{heath_enhancing16} of a random subset of antennas.
This produces randomized radiation patterns on
side lobes, with minimal impact on normal transmissions\footnote{
Due to space limit, we omit details about this 
defense. We refer interested readers to related work for more information.}.

Antenna-based defenses require TX to change the selected antenna subset very
frequently, often on a per-symbol basis, \ie
at the time scale of {\em sub-nanoseconds}.
Less frequent switching keeps signals within a packet 
highly correlated with each other. This could allow the attacker
to simply estimate the wireless channel, or guess the correlation constant
to recover the transmission.
Despite the effectiveness, switching at a per-symbol frequency 
incurs extremely high cost in hardware and power. 
Today's hardware can only support packet-level switching (10s of nanoseconds)
for the same reason, making antenna-based defenses impractical.

Despite the impracticality, we implemented the defenses in simulation. 
We found it effectively defends against single-device side-lobe 
eavesdroppers, regardless of where the attack is launched.
However, it remains vulnerable to advanced attacks. 
For instance, attack can use multiple \textit{synchronized}
devices to measure
side-lobe signals at different angles, 
undo the effects of antenna randomization on 
a per-symbol basis, and recover the packets.
\yedit{The key is to decode the antenna selections for transmission
from measurements, as there is a limited number of antenna subset selections.}

\para{RF-Chain-Based Defenses.}
Unlike antenna-based defenses,
these defenses add \textit{additional} RF chains 
to generate noise and do not need randomizations in TX's radiation pattern. 
They ``jam'' the eavesdropper at TX's side lobes,
so the attacker can only receive a mixture of transmitted signals
and noise signals.
For mmWave hardware, this adds significant complexity
and cost in RF signal processing components, increasing the
hardware cost and power requirements.
Despite that previous work~\cite{rfchain_high_power,
heath_enhancing16} reduces the hardware requirement, 
these defenses~\cite{heath_enhancing16, 
juying17, mckay13, heath_adhoc16} remain costly and 
power-demanding.

We found in simulations that RF-chain-based defenses effectively
defend against single-device eavesdroppers. 
Although, TX's side lobes have gaps in between, 
which nulls the transmitted signals. An advanced attacker
can exploit this and search for only noise signals.
He could then perform noise cancellation with only two synchronized receivers: 
one listening to only noise 
and the other eavesdropping the mixed noise and legit signals.
The attack becomes more difficult when 
TX uses over two RF chains to generate noise. 
Noise from different RF chains would mix together and becomes
difficult to isolate. Still, this countermeasure comes at an
even higher cost in mmWave hardware and device power.

\begin{table}[h]
\centering
\resizebox{0.95\columnwidth}{!}{%
\begin{tabular}{|l|c|c|c|c|}
\hline
\multirow{2}{*}{\begin{tabular}[c]{@{}c@{}}Category\end{tabular}} 
  & 
\multicolumn{2}{c|}{Defense Requirement} & 
\multicolumn{2}{c|}{Vulnerability} \\ 
\cline{2-5}
   & 
\begin{tabular}[c]{@{}c@{}}\# of RF\\Chains\end{tabular} & 
\begin{tabular}[c]{@{}c@{}}Antenna\\Switching\\Frequency\end{tabular} & 
\begin{tabular}[c]{@{}c@{}}\# of Sync.\\Devices\\to Attack\end{tabular} & 
\begin{tabular}[c]{@{}c@{}}Info. Required \\ for Attack\end{tabular} \\ 
\hline
No Defense & 1 & - & 1  & side-lobe signals \\ 
\hline
Antenna-Based & 1 & per-symbol & N & \begin{tabular}[c]{@{}c@{}}signals at\\N locations\end{tabular} \\ 
\hline
RF-Chain-Based & $>$2 & - & 2 & \begin{tabular}[c]{@{}c@{}}noise signals \\ at N locations\end{tabular} \\ 
\hline
\end{tabular}
}
\vspace{-0.05in}
\caption{Summary and vulnerabilities of different defense mechanisms.
$N$ is the number of TX antennas.}
\label{table:defense}
\vspace{-0.0in}
\end{table}

\vspace{-0.08in}
\section{Related Work}
\label{sec:related}

\para{Security Analysis in mmWave Eavesdropping.}
Existing works to study mmWave eavesdropping either perform 
simulations~\cite{alotaibi_tcom16, 
dai2011exploring, dai2013eavesdropping, heath_enhancing16, juying17, 
kim2017analysis, heath_tcom13, wang_twcom16, heath_adhoc16}
or use horn antennas~\cite{steinmetzer2015eavesdropping}, which
have no side lobes. 
Differing from these, 
we are the first to study mmWave side-lobe eavesdropping 
from actual measurements, using commercial 60GHz phased arrays
in real-world application scenarios. 

Many of these proposed defenses against mmWave eavesdropping, \ie 
using antenna-based~\cite{alotaibi_tcom16, 
heath_enhancing16, heath_tcom13} 
or RF-chain-based designs~\cite{
heath_enhancing16, juying17, wang_twcom16, heath_adhoc16} 
assume a naive single-device attacker. 
Our work analyzes these proposals and finds these methods 
either as vulnerable to advanced attackers
with multiple synchronized devices, or they introduce significant 
hardware overhead and cost. Thus, these defenses are not applicable to
mmWave transmissions.

\para{Eavesdropping in Low-Frequency RF Bands.}
Eavesdropping is more prevalent and easier in lower frequency 
bands, \eg Wi-Fi and cellular, due to its omni-directional signals. 
Many previous works propose defense mechanisms using jamming, which injects
artificial noise towards the 
attackers~\cite{cooperative_jam09, relay_jamming08, 
mckay13}. Although different techniques are used, \eg a separated jammer 
synchronized with the transmitter~\cite{ijam10}, cooperative devices 
or relays~\cite{cooperative_jam09, relay_jamming08}, these defensive 
mechanisms all require a high number of RF chains. Despite the acceptably 
minimized hardware cost in commodity Wi-Fi and cellular
devices, the cost of these defenses remains extremely high in the context of mmWave.

\section{Conclusion and Future Work}
\label{sec:discussion}

Despite an initial step to investigate mmWave side-lobe
eavesdropping with real measurements, we already find it proves to be
a much greater threat than we expected. 
We hope our results
draw the attention of the community and shed light on the future 
development of mmWave communications. 
Moving forward, many open challenges remain. 

\para{Potential Defenses.} Despite existing proposals,
we lack a practical and secure solution against side-lobe
eavesdropping. Other than reducing the RF chain cost in mmWave
communications, a possible alternative could leverage
the antenna artifacts. Designing specific artifacts in hardware
could resist the attack since we saw earlier that artifacts
may alter the shape of side lobes. The artifacts should be carefully
designed so the normal transmission remains unaffected.

\para{Empirical Validation of Advanced Attacks.} 
We briefly described and simulated two types of advanced attacks, 
\ie antenna randomization attack and noise cancellation attack. 
While other advanced attacks remain possible, 
current mmWave hardware is not flexible enough to implement these attacks.
Also, our device does not report bit error rate (BER) which may shed
light on more fine-grained insights as~\cite{imc10mesh} did.
We hope more flexible hardware becomes available soon, so we can
empirically validate the attacks with consideration of antenna artifacts, 
which may affect the attacks' performance.

\bibliographystyle{acm}
\balance
\begin{small}
\balance
\bibliography{refs,ref_heather,ref_madhow_pubs,ref_zheng,ref_physec}
\end{small}

\end{document}